\documentclass[aps,preprint,groupedaddress]{revtex4b1}
\usepackage{epsfig}

\begin{document}
\bibliographystyle{apsrev}

\title{Vortex Flow and Transverse Flux Screening
		at the \\ Bose Glass Transition}

\author{A.\ W.\ Smith, H.\ M.\ Jaeger and T.\ F.\ Rosenbaum}

\affiliation{The James Franck Institute and Department of Physics,  The
		University of Chicago,\\ Chicago, IL 60637 }

\author{ A.\ M.\ Petrean, W.\ K.\ Kwok and G.\ W.\ Crabtree}

\affiliation{Materials Science Division, Argonne National Laboratory, 
		Argonne, IL 60439 }


\begin{abstract}
We investigate the vortex phase diagram in untwinned
YBa$_2$Cu$_3$O$_{7-\delta}$ single crystals with columnar defects.
These randomly distributed defects, produced by heavy ion irradiation,
are expected to induce a ``Bose Glass'' phase of localized vortices
characterized by a vanishing resistance and a Meissner effect for
magnetic fields transverse to the defect axis. We directly observe the
transverse Meissner effect using an array of Hall probe
magnetometers. As predicted, the Meissner state breaks down at
temperatures $T_{s}$ that decrease linearly with increasing transverse
magnetic field. However, $T_s$ falls well below the conventional
melting temperature $T_m$ determined by a vanishing resistivity,
suggesting an intermediate regime where flux lines are effectively
localized even when rotated off the columnar defects.
\end{abstract}


\maketitle

Weak pinning, large anisotropy, and prodigious thermal fluctuations in
high temperature superconductors permit magnetic vortices to exist in
a liquid phase which flows easily in response to an external
current~\cite{1}.  This technologically undesirable state of finite
electrical resistance is only quelled at sufficiently low temperatures
when vortex-vortex interactions precipitate a solid phase~\cite{2,3}.
Columnar defects, a form of strong disorder correlated across many
CuO$_2$ planes, serve as an effective strategy to bolster the
constraining effects of the interaction potential, localizing the
vortices on the disordered array of columns.  The resulting ``Bose
Glass" phase~\cite{4} exhibits two characteristic signatures: a
diverging vortex viscosity leading to a zero resistance state, as well
as a diverging tilt modulus resulting in vortex alignment with the
columns even as the external magnetic field rotates. A large rotation
of the external field, however, will melt the glass at fixed
temperature. We monitor both the onset of a resistive response (vortex
mobility) and the onset of transverse flux penetration (vortex
alignment). In contrast to theory, we find that these two events yield
widely different demarcations for the Bose glass phase.

Amorphous columnar tracks created by heavy ion bombardment of cuprate
superconductors are of nearly optimal size and geometry to pin
vortices over their entire length.  Thermal excitations, however,
permit some segment of a vortex line to wander off its pinning
site. Nelson and Vinokur~\cite{4} have mapped this statistical
mechanics problem onto the quantum mechanical problem of
two-dimensional bosons in the presence of point disorder~\cite{5}.
They find a sharp phase transition between a high-temperature vortex
liquid of delocalized and entangled vortex lines and a low temperature
Bose glass dominated by the spatial disorder in the columnar defect
distribution. The predictions of scaling theory~\cite{4,6,7} have been
tested~\cite{8,9,10,11,12,13,14,15,16,17} by measuring the vortex
mobility subjected to a driving current at the approach to the glass
transition temperature, $T_{BG}$. These experiments vary widely on
values for the critical exponenets and on the angular dependence of
$T_{BG}$. We consider an additional prediction of the theory, that
below $T_{BG}(\vec{H})$ vortices remain parallel to the columnar
defects even when the external magnetic field is rotated off the
defect axis.  While this is a striking element of the theory, it has
so far been untapped as a quantitative probe of the nature of the
glass transition~\cite{18}.

With our experimental configuration, illustrated in
Fig.~\ref{fig:Setup}, it is possible to measure \emph{both} the
vortex mobility and the extent to which flux lines remain parallel to
the columnar defects. We sense the vortex mobility using an 8 turn,
0.3 mm diameter pickup coil near the sample. Since each vortex carries
a fixed amount of magnetic flux, changes in the vortex density appear
as an induced voltage in the pickup coil.  In the presence of a small
ac driving field (less than 0.2 gauss and at 1000 Hz), this
voltage signal provides a measure of how easily vortices hop from
columnar defect to columnar defect.  A micron-sized InAs Hall probe
array~\cite{19} under the sample measures the local magnetic field
component perpendicular to the columnar defects.  This provides a
direct gauge of vortex rotation in response to the transverse
component of the applied field.

The YBa$_2$Cu$_3$O$_{7-\delta}$ untwinned single crystal with
dimensions 0.5$\times$0.5$\times$0.035 mm$^3$, critical temperature
$T_c = 92.5$ K, and transition width $\Delta T_c=0.3$ K was irradiated
with 700 MeV Sn ions at Argonne's ATLAS source, producing columnar
defects parallel to the c-axis with equivalent vortex density
$B_{\Phi}$ = 0.5T.

We cooled the YBa$_2$Cu$_3$O$_{7-\delta}$ single crystal from the
normal state at a fixed parallel magnetic field $H_{\parallel}$ and
then ramped up a perpendicular field $H_{\perp}$. The parallel
component of the field was aligned carefully with the columnar defects
using a 3-axis split-bore magnet system, with the magnetic field
direction adjusted to minimize the vortex mobility determined by the
pick-up coil. The Hall probes recorded the local perpendicular
magnetic field during each ramp cycle. The local field underneath the
crystal remained substantially reduced from that on either side at
small $H_{\perp}$ (Fig.~\ref{fig:Setup}), indicating a 30\%
diamagnetic shielding of the external field~\cite{20}.  This behavior
is independent of the direction of the field ramp, ruling out an
alternative explanation of vortex penetration in a Bean profile.  At
sufficiently large $H_{\perp}$, hysteresis sets in, the field profile
acquires a component that depends on ramp direction and the vortices
no longer remain aligned with the columns. In order to demonstrate
this effect, we consider the perpendicular magnetization $M_{\perp}$
taken as the difference between the perpendicular magnetic field
$B_{\perp}$ underneath the sample with that far
away. Figure~\ref{fig:Magnetization} shows $M_{\perp}$ as the applied
field $H_{\perp}$ was swept through cycles $H^{max}_{\perp}\rightarrow
- H^{max}_{\perp}\rightarrow H^{max}_{\perp}$ with successively
increasing ramp end points $\pm H^{max}_{\perp}$. For small
$H_{\perp}^{max}$, $M_{\perp} \propto - H_{\perp}$, and there is
negligible hysteresis; together these two results indicate that the
superconductor expels the component of applied field perpendicular to
the columnar defects.  At sufficiently large $H_{\perp}^{max}$,
$M_{\perp}$ becomes hysteretic and is no longer proportional to
-$H_{\perp}$. For larger $H_{\perp}^{max}$, $M_{\perp}$ decreases even
further relative to the dashed line indicating that the magnetic field
direction tracks that of the applied field H, and the transverse
Meissner effect is lost.

We quantify this breakdown of the transverse Meissner effect in terms
of a critical tilt angle
$\theta_c=tan^{-1}(H_{\perp}^c/H_{\parallel})$.  In order to determine
$H^c_{\perp}$, the field at which transverse flux enters the
superconductor, we measured the hysteresis in the local field under
the sample $B_{\perp}$ as a function of $H_{\perp}^{max}$.  We plot in
the inset to Fig.~\ref{fig:Screen} the hysteresis width of the
resulting magnetization loops $B_{\perp}(H_{\perp})$. This width is
seen to remain negligible until $H_{\perp}$ reaches a critical value
$H_{\perp}^c$, beyond which it increases sharply.  We have checked
explicitly that the value for $H^c_{\perp}$ obtained in this way does
not depend on position underneath the the crystal; it is the same for
Hall probes near the sample center (\#4) and near the sample edge
(\#6). By contrast, the hysteresis width measured in an unirradiated
crystal at the same temperature and parallel magnetic field in an
unirradiated YBCO crystal shows no such offset in $H_{\perp}$.

Our direct observation of a transverse Meissner effect validates a
central prediction of the Bose glass theory.  Moreover, from
measurements of the critical transverse field $H_{\perp}^c(T)$ over a
range of temperatures we can define a maximum screening temperature
$T_s(H_{\perp})$ above which the transverse Meissner effect breaks
down.  The main panel in Fig.~\ref{fig:Screen} shows $T_s$ as a
function of tilt angle. With increasing $B_{\parallel}$ the ability to
screen transverse flux diminishes and smaller tilt angles are required
to significantly suppress $T_s$.  Specifically, for applied fields $H
= 0.4 B_{\Phi}$ and $1.0B_{\Phi}$ our data over a range of more than
40K exhibit a linear decrease of $T_s$ with $\theta$.  Here $B_{\Phi}$
is the matching field at which the vortex density equals the density
of columnar defects~\cite{21}. Once $H_\parallel>B_{\Phi}$ screening
appears to become rapidly ineffective; at
$H_{\parallel}$=2.6$B_{\Phi}$, $T_s$ has been suppressed by 50\%
within the first $0.1\deg$ of tilt.

We now compare these results for the transverse Meissner effect with
measurements of the onset of vortex mobility.  Using the pickup coil
shown in Fig.~\ref{fig:Setup}, we tracked the locus of temperature and
magnetic field data for which the linear resistivity $\rho$ drops
below a fixed threshold (taken here as 1/1000 of the normal state
resistivity at current density $j \simeq 10$ A/cm$^2$).  Given
the strong temperature dependence of the resistivity $\rho(T)$ as the
zero-resistance state is approached, this type of resistance threshold
criterion conventionally has been used to monitor the melting
temperature $T_m$.  The normalized change in the melting temperature,
$\Delta T_m/T_m=[T_m(H_{\perp}=0)-T_m(H_{\perp})]/T_m$, reveals a cusp
over a broad range of tilt angles $\theta$ (Fig.~\ref{fig:Coil}a).  By
comparison the screening temperature $T_s$ in Fig.~\ref{fig:Screen},
the melting temperature $T_m$ varies non-linearly and only weakly with
$H_{\perp}$.

If we associate the Bose glass temperature $T_{BG}(H_{\perp})$ with
the demise of the transverse Meissner effect by setting
$T_s(H_{\perp}) = T_{BG}(H_{\perp})$, we can determine the static
critical exponent $\nu$ for the glass transition via the
relation~\cite{7} $T_{BG}(H_{\perp}\! =\! 0) -T_{BG}(H_{\perp})
\propto H_{\perp}^{1/\nu}$.  From the data for $H_{\parallel} =
0.4B_{\Phi}$ and 1.0$B_{\Phi}$ (Fig.~\ref{fig:Screen}), we find $\nu=
0.9 \pm 0.1$ and $1.0 \pm 0.1$, respectively.  This result is at the
bound~\cite{22,23} $\nu\! =\! 2/d =1$, with effective system
dimensionality d = 2, and it agrees with computer simulations~\cite{7}
of Bose glass melting.  Our data indicate that even well above
$B_{\Phi}$ the exponent $\nu$ remains approximately 1.
As shown in Fig.~\ref{fig:Coil}b, $\Delta T_m/T_m$ over two decades
follows a power law in tan$\theta$ or, equivalently, in $H_{\perp}$.
Taking the melting temperature $T_m$ as the Bose glass temperature
would lead to a critical exponent $\nu = 1.9 \pm 0.2$ for $\theta >
1$, but with rounding of the cusp at the smallest tilt
angles~\cite{24}. This value of $\nu$ is consistent with that reported
in Ref.~25 where $T_m(H_{\perp})$ was determined from transport
measurements.

The differences between $T_m$ and $T_s$ also lead to a completely
different demarcation between the glass and liquid phases.  We combine
in Fig.~\ref{fig:Phase} the phase diagrams based on vortex flow and
transverse field screening for $H_{\parallel} = 0.4B_{\Phi}$.  The
boundary between Bose glass and vortex liquid coincides for both types
of measurements at $H_{\perp}=0$, but $T_m$ and $T_s$ quickly
bifurcate with increasing transverse field.  Not only is the
transverse screening limited to a far narrower range of magnetic field
tilt angle, it also indicates a Bose glass transition at temperatures
many tens of Kelvin below the disappearance of the linear resistive
response.

We are left with the intriguing scenario where a new, intermediate
regime exists between the glass and the liquid state which exhibits no
vortex flow even in the absence of a fully developed transverse
Meissner effect.  The origin and detailed physical nature of this
intermediate regime between $T_m$ and $T_s$ remain open questions.
Our results appear to be in conflict with the current theoretical
framework for two-dimensional bosonic states which allows for only a
single transition from insulating to superconducting behavior.
However, there has been mounting experimental evidence in different
systems~\cite{26} that an intermediate regime may indeed exist.
Comparative studies of high-temperature superconductors with different
types of disorder and different degrees of anisotropy, as well as
characterization of the full non-linear spectral response, should help
illuminate this unusual vortex state.

\begin{acknowledgments}
We thank Dr.\ Reggie Ronningen for assisting with the
ion irradiation at the National Superconducting Cyclotron Laboratory.
This work was supported by the National Science Foundation through the
Science and Technology Center for Superconductivity (DMR91- 20000) and
under contract \#PHY-9528844, and by the U.S. Department of Energy,
BES-Materials Science under contract \#W-31-109-ENG-38 (WKK, AMP,
GWC).
\end{acknowledgments}


\centerline{\epsfig{figure=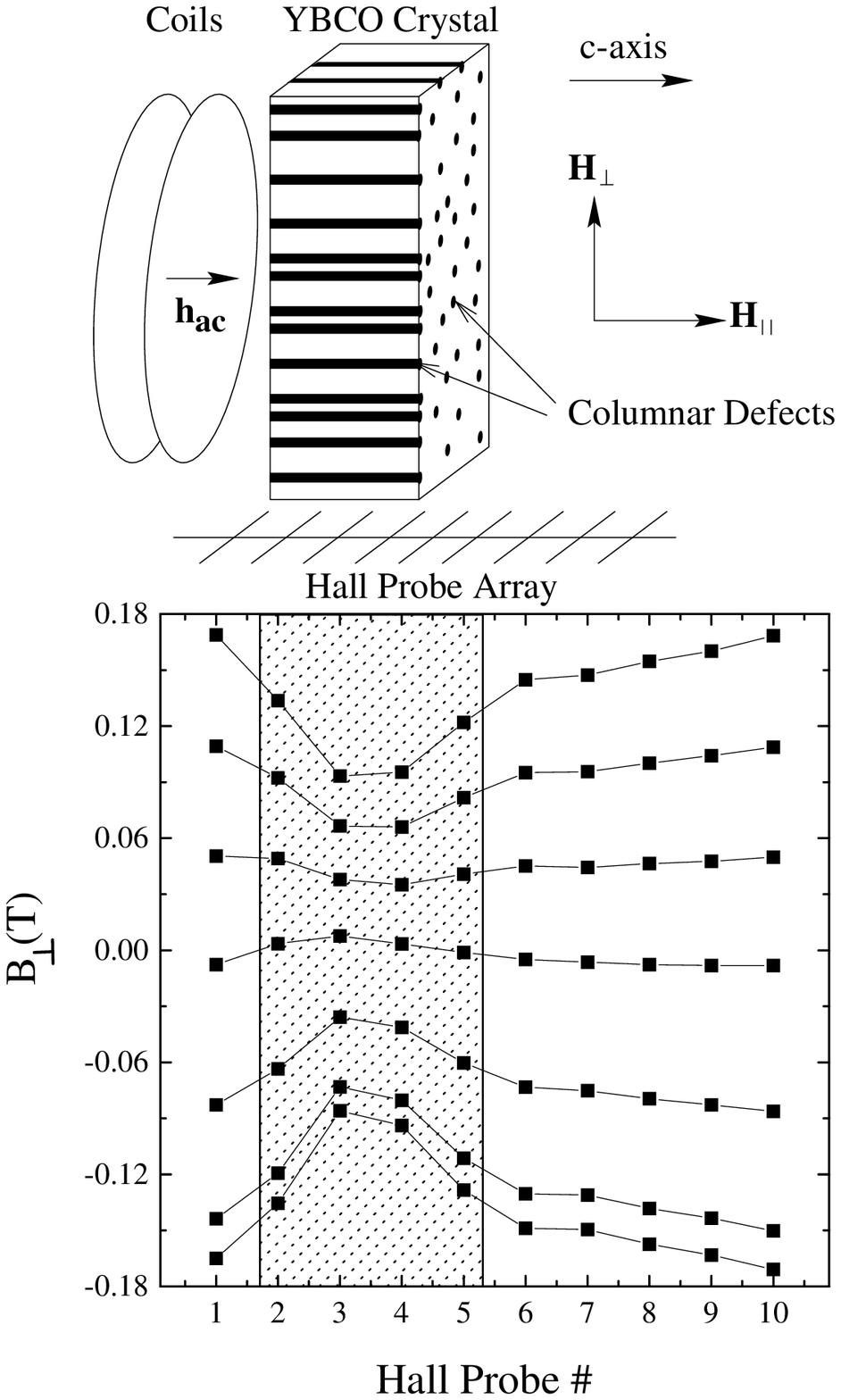,height=9in}}
 
\centerline{\epsfig{figure=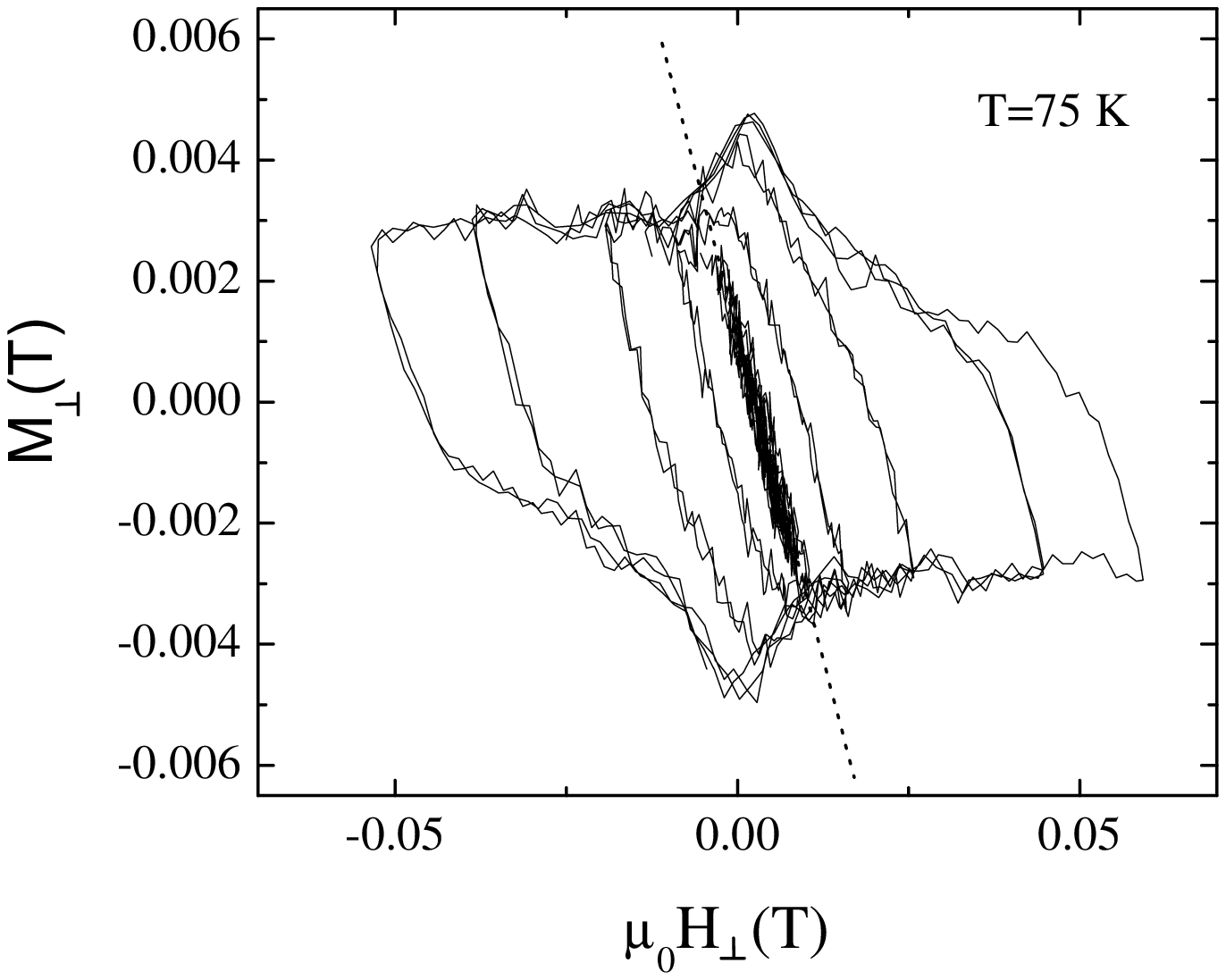,height=9in}}

\centerline{\epsfig{figure=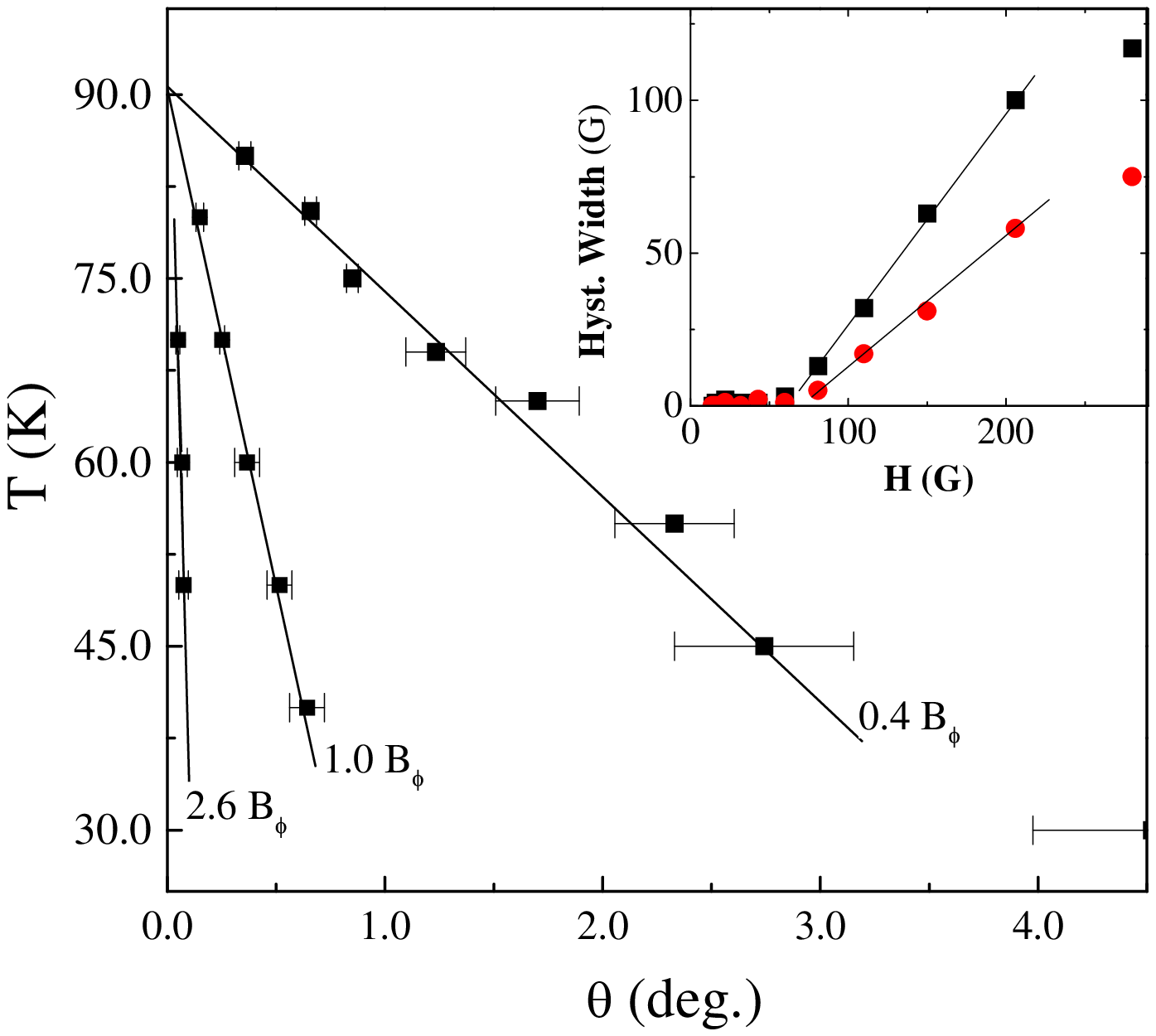,width=7in}}

\centerline{\epsfig{figure=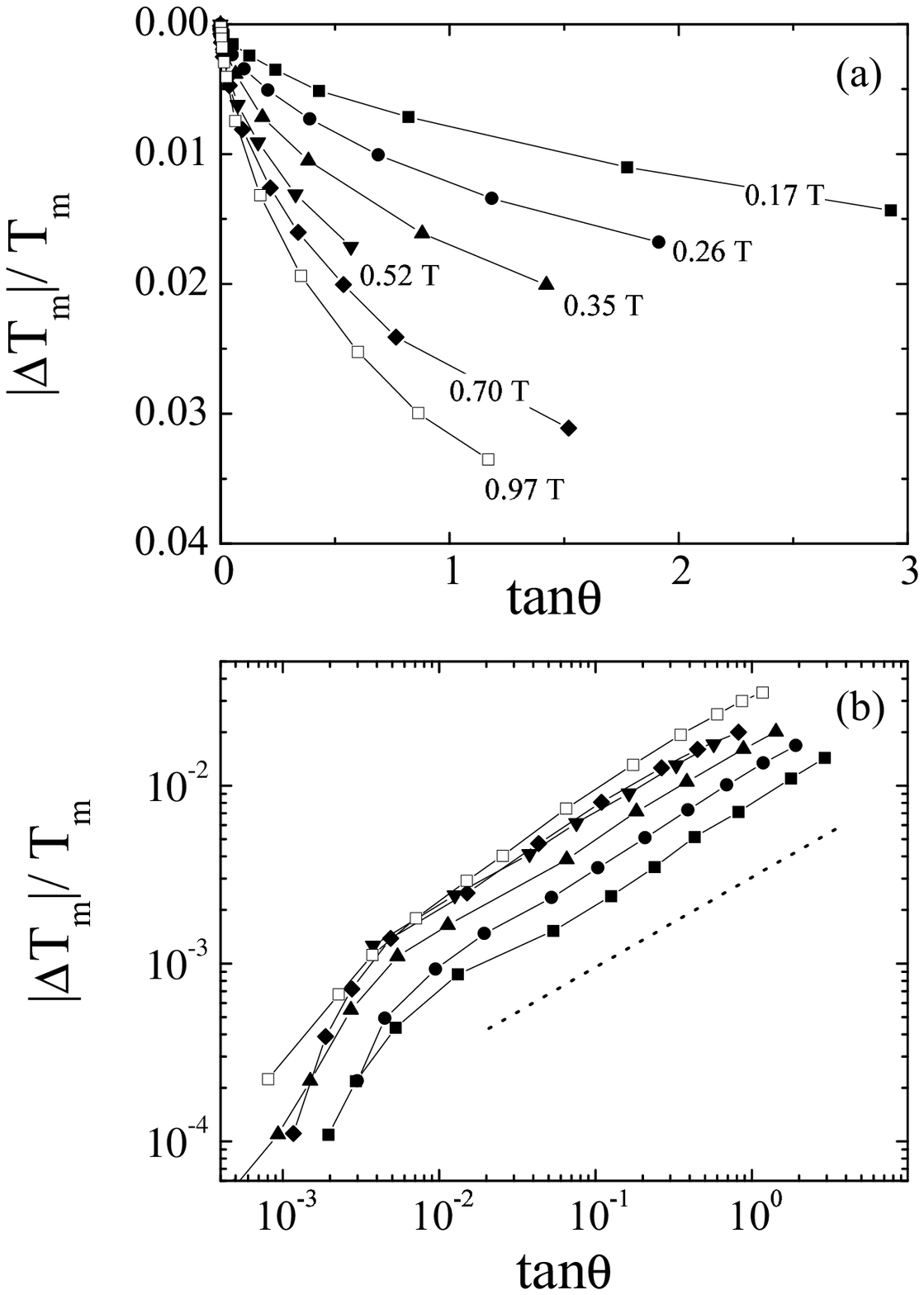,height=9in}}

\centerline{\epsfig{figure=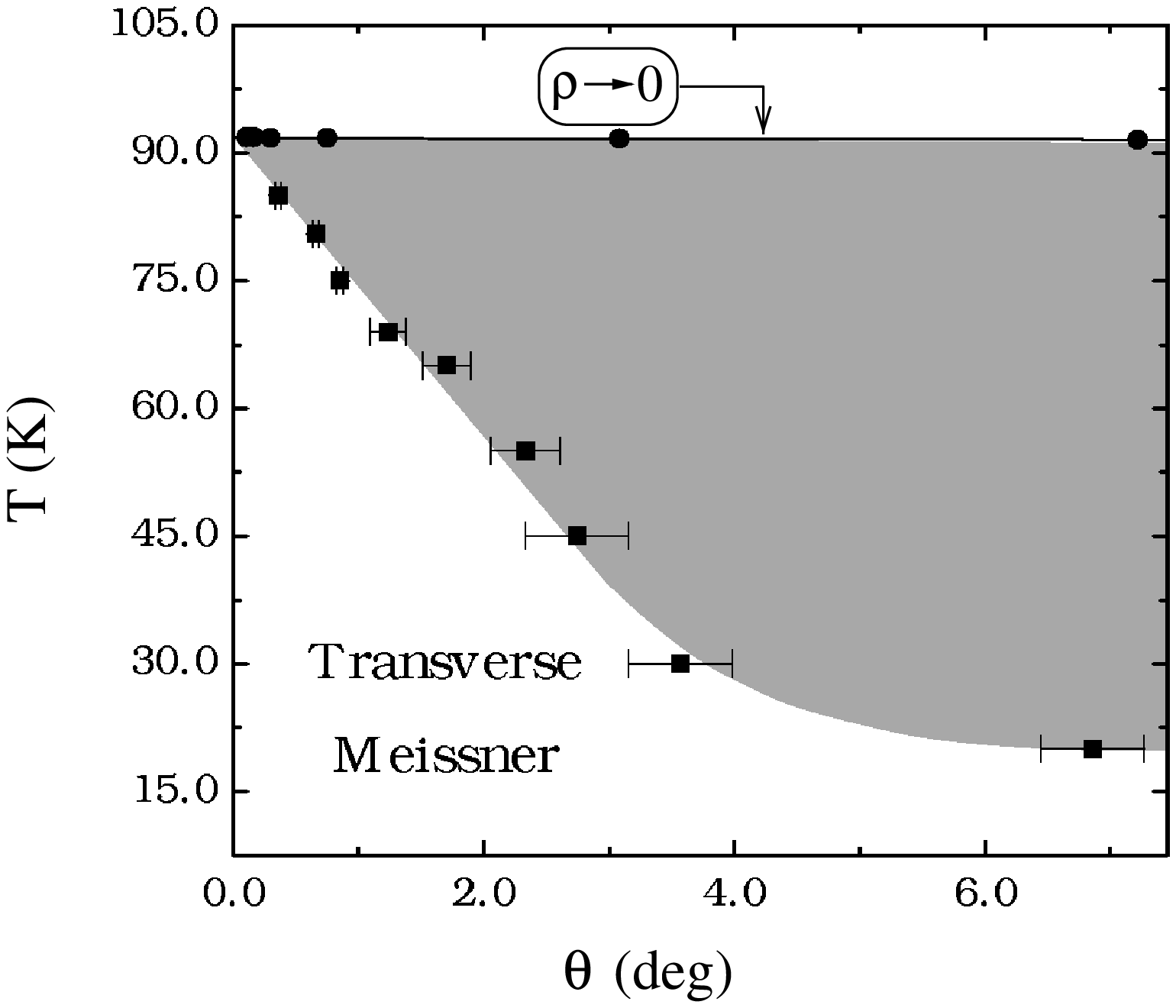,height=9in}}

\clearpage


\begin{thebibliography}{}
\bibitem{20} We do not expect the field under the sample to remain
unchanged during the field ramp even when screening is perfect because
there is always a non-zero distance between the Hall probe array and
the sample.
\bibitem{24} Taking the recent correction (Refs.~6,7) to the scaling
function into account, we note that previous experimental
determinations of $\nu$ from vortex flow measurements in YBCO samples
have found values of $\nu$ ranging from 1 to 3 (Refs.~9,12,14,17,18).
Our vortex flow data range extends to much smaller tilt angles, and
within the scatter the data for $\theta < 1^o$ are consistent with
$\nu \simeq 1$.  Whether $\nu$ = 1 is the true static critical
exponent rather than $\nu$ = 1.9 cannot be stated unambiguously from
our data for $\theta < 1^o$ because such a fit is over a limited range
and simply may be due to an intrinsic splay of heavy ion tracks about
the crystal's c-axis .
\end{thebibliography}
\end{document}